\documentclass[%
 reprint,
 amsmath,amssymb,
 aps,
]{revtex4-2}
\usepackage{xcolor}%
\usepackage{graphicx}
\usepackage{dcolumn}
\usepackage{bm}
\usepackage{mathtools} 
\usepackage{stmaryrd}

\newcommand{\Ivec}{\mathbf{I}}
\newcommand{\Qvec}{\mathbf{Q}}

\graphicspath{{Pictures/}} 

\begin{document}

\preprint{APS/123-QED}

\title{A Simple Tensorial Theory of Smectic C Liquid Crystals}
\author{Jingmin Xia}
\email{jingmin.xia@nudt.edu.cn}
\affiliation{College of Meteorology and Oceanography, National University of Defense Technology, China}
\author{Yucen Han}
\email{yucen.han@strath.ac.uk}
\affiliation{Department of Mathematics and Statistics, University of Strathclyde, Glasgow, UK}
\begin{abstract}
The smectic C (smC) phase represents a unique class of liquid crystal phases characterised by the layered arrangement of molecules with \emph{tilted} orientations with respect to layer normals.
Building upon the real-valued tensorial smectic A (smA) model in [Xia et al., PRL, 126, 177801 (2021)], we propose a new continuum mathematical model for smC (and smA) by introducing a novel coupling term between the real tensor containing orientational information and density variation, to control the tilt angle between directors and the layer normal (the tilt angle is zero for smA and nonzero for smC).
To validate our proposed model, we conduct a series of two- and three-dimensional numerical experiments that account for typical structures in smectics: chevron patterns, defects, dislocations and toroidal focal conic domains (TFCDs). These results also reveal the phenomenological differences between smA and smC configurations.
\end{abstract}
\maketitle

\begin{figure}   
\includegraphics[width=0.7\columnwidth]{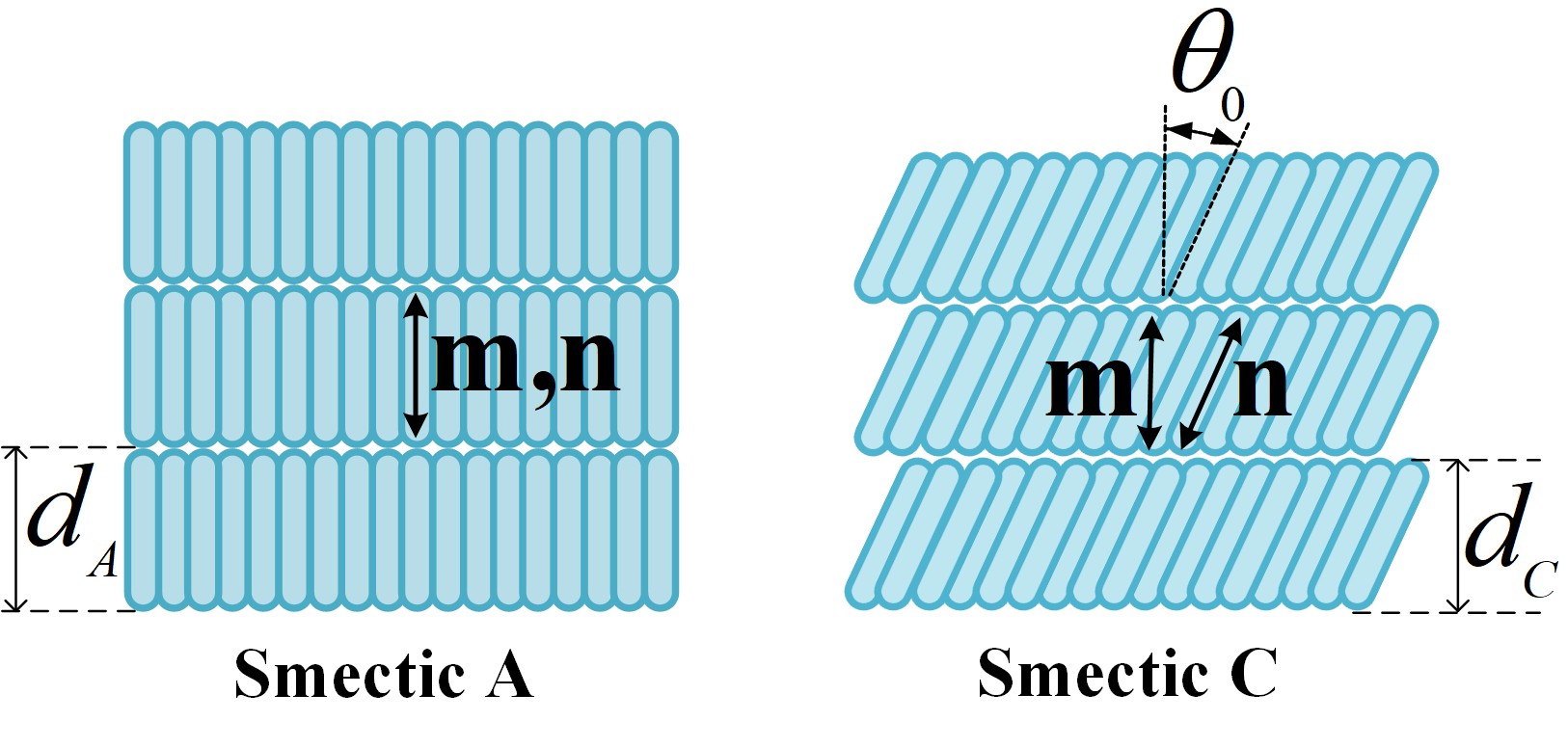}
    \caption{A schematic illustration of smA and smC phases. In smA, the orientation of LC molecules $\mathbf{n}$ is well-defined; they align along the layer normal $\mathbf{m}$, resulting in a layer structure. In smC, the orientation of molecules $\mathbf{n}$ deviates from the layer normal $\mathbf{m}$ with a tilt angle $\theta_0$. Blue rods represent LC molecules. $d_A$ and $d_C$ are the layer thickness of smA and smC, respectively.
    }
    \label{fig:smAC}
\end{figure}

\textit{Introduction}--Liquid crystals (LCs), an intermediate state of matter between the ordered solid state and the disordered fluid state, have been a subject of profound interest in both the scientific community and the technological world \cite{gennes-book, stewart-2004-book}.
Among various LC phases, smectic phases stand out for their layer-like arrangement of molecules, featuring periodic density undulation \cite{zappone-2022-article}.
Two of the most prominent smectic phases are the smectic A (smA) and smectic C (smC, also known as tilted smectic) phases, each characterised by distinct molecular orientations within the layers. A schematic illustration on these two phases can be seen in Fig.~\ref{fig:smAC}. 

Researchers studied various smA and smC systems in experiments and exploited their special properties to develop potential applications.
For smA and smC in shells and droplets,
complex three-dimensional (3D) chevron structures and numerous macroscopic defects called focal conic domains (FCDs) are observed with distinct optical textures \cite{urbanski2017liquid}. 
The formation of depression during the local smA-smC transition allows researchers to use smectics as an active substrate to manipulate nano-particles 
\cite{gharbi2021liquid}.
Ferroelectric displays, which require a smC host utilising faster switching modes, may bring about innovation in the display industry \cite{mandle2015control}.
Furthermore, the smectic-based sensors have an outstanding sensitivity compared to nematic-based sensors 
in response to a subtle change in the chain rigidity of amphiphiles \cite{chen2022dissipative}.

In order to deeply understand the macroscopic phenomena of smA and smC LCs, a continuum mathematical model that can describe smA and smC phases under the same framework is in demand. Although there have been some successful continuum models for smA in recent years \cite{ball2023free, paget-2023-article}, these models are difficult to extend to smC phases.
So far, there are a few continuum models for smC LCs.
The Orsay Group were the first to present a continuum model for the smC phase in the 1970s, assuming small perturbations of planar-aligned samples of smC \cite{orsay1971simplified}.
With the phenomenological approach, the system of smC with fixed layer normal was studied by analogy to the behaviour of superconductors with a complex scalar field in \cite{gennes-1973-article}. With mean-field theory, a Landau--Ginzburg free energy was developed in \cite{chen-1976-article}. Leslie and Stewart et al.'s work \cite{leslie-1991-article} introduced the tilt vector, the unit orthogonal projection of the director onto the local smectic planes, and the information of tilt angle between the layer normal and the director was hidden in the elastic constants. The elastic energies for smC were further extended to describe layer dilation and moderate variations of tilt angle by \cite{blake1996original}. Later, \cite{biscari-2007-article} introduced the $\Qvec$-tensor from Landau--de Gennes  theory \cite{gennes-book} for describing directors into the smC model and 
the corresponding free energy included two competing coupling terms, which prefer the layer normal to be parallel or orthogonal to the director.
However, due to the limitations 
of the above smC models, research on modelling smC LCs 
mostly stays at the analytical level, and numerical simulations for complex structures are rarely possible, let alone research on frustrated smC systems in confinement. Another problem is that the tilt angle between the director and the layer normal is not directly quantified within the models. The explicit description of the tilt angle in smC can provide more precise guidance for the application of smA and smC liquid crystals.

In this letter, we deal with the above problems by developing a new continuum model for smC (and smA), 
with a real-valued order parameter $\delta\rho$, representing the variation from the average density, and a tensor-valued order parameter $\mathbf{Q}$ from the LdG theory. The tilt angle $\theta_0$ is explicitly included in our proposed model by modifying the coupling term depicted in Xia et al.'s smA model \cite{xia-2021-article}. Within this new model, typical smectic structures, like bookshelf and chevron \cite{biscari-2007-article}, and dislocation \cite{meyer-1978-article} in smA and smC are numerically recovered. We investigate the different behaviour of point defects in two-dimensional (2D) smA and smC, and numerically deduce that half-integer defects do not exist in the smC phase.
The model is also validated by obtaining toroidal focal conic domains (TFCDs) \cite{perez-1978-article} confined in a 3D box.

\textit{Mathematical Model}--To describe the smC (and smA) phase, we need two order parameters, one for the layering structure and the other representing the orientation of LC molecules. 
According to the discussion in \cite{xia-2021-article}, a real-valued order parameter $\delta\rho$ and a tensor-valued order parameter $\mathbf{Q}$ are chosen.
The $\delta\rho$ represents the variation from the average density. The $\mathbf{Q}$ contains the information of the preferred directions of spatially averaged local molecular alignment or the directors, and the corresponding orientational orders.

Based on the mathematical model in \cite{xia-2021-article}, the energy of a smC (and smA) system can be written as a sum of a Landau--de Gennes (LdG) \cite{gennes-book} term of $\mathbf{Q}$, a term of $\delta\rho$ to describe layer structure, and a coupling term between $\delta\rho$ and $\mathbf{Q}$ to control the tilt angle $\theta_0$ between the director and the layer normal (see Fig.~\ref{fig:smAC}) on domain $\Omega$,
\begin{equation}\label{eq:Functional}
    F(\delta\rho,\mathbf{Q}) = \int_{\Omega} 
     \left\{f_{LdG}(\Qvec) + f_{sm}(\delta\rho) + f_{tilt}(\delta\rho,\Qvec)\right\}.
\end{equation}
(i) The simplest LdG free energy density is given by
\begin{equation}
    f_{LdG}(\Qvec) = f_b(\mathbf{Q}) + f_{el}(\nabla\mathbf{Q}),
\end{equation}
which includes the bulk energy density
\begin{equation}
\label{eq:ldg-density}
    f_b(\mathbf{Q}) = -\frac{A}{2} \text{tr}(\Qvec^2) - \frac{B}{3} \text{tr}(\Qvec^3) + \frac{C}{2} \left(\text{tr}(\Qvec^2)\right)^2,
\end{equation}
and the elastic energy density 
\begin{equation}\label{eq:el}
    f_{el}(\nabla\Qvec) = \frac{K}{2}|\nabla\Qvec|^2.
\end{equation}
The parameter $A$ in \eqref{eq:ldg-density}, is a reduced temperature; the parameter $B$ is positive (resp.\ negative) for rod-like (resp.\ disc-like) LC molecules; and the parameter $C$ is taken to be positive so to keep the functional bounded from below.
Here, $\mathrm{tr}(\cdot)$ represents the trace operator. Note that the global minimiser of the bulk energy $f_{b}$ is a uniaxial $\Qvec$ \cite{majumdar-2010-article}, i.e., $\Qvec$ can be formulated as $\Qvec=s(\mathbf{n}\otimes \mathbf{n}-\frac{\Ivec_d}{d})$, where $\mathbf{n}$ is an arbitrary unit vector in $\mathcal{S}^{d-1}$, $d$ is the spatial dimension. The orientational order $s$ can be tuned by varying parameters $A$, $B$ and $C$. This LdG energy term mainly characterises the isotropic-nematic phase transition. The reported values for MBBA as a representative nematic material are $B = 0.64\times 10^4 N/m^2$ and $C = 0.35\times 10^4 N/m^2$ \cite{wojtowicz1975introduction}.
To ensure the imposition of the orientational order $s = 1$ for minimisers of $f_b$ (one can check this by simple calculations or using \cite[Proposition 15]{majumdar-2010-article}), we take $f_{b} = -\frac{l}{2} \left(\text{tr}(\Qvec^2)\right) - \frac{l}{3} \left(\text{tr}(\Qvec^3)\right) + \frac{l}{2} \left(\text{tr}(\Qvec^2)\right)^2$ in 3D and $f_{b}=-l \left(\text{tr}(\Qvec^2)\right) + l \left(\text{tr}(\Qvec^2)\right)^2$ in 2D (also known as the reduced Landau--de Gennes model in \cite{han-2020-reduced}). 
The $f_{el}$ in \eqref{eq:el} is the simplest form of the elastic energy density under the one-constant approximation that penalises spatial inhomogeneities, and $K$ is the material-dependent elastic constant. The typical value of the elastic constant for MBBA is $K = 4\times 10^{-11}N$ \cite{wojtowicz1975introduction}. We also note that it is feasible to apply the anisotropic elastic form for \eqref{eq:el}.\\
(ii) The smectic potential $f_{sm}$ drives the appearance of smectic layers,
\begin{equation}
    f_{sm}(\delta\rho) = \frac{a}{2}(\delta\rho)^2+\frac{b}{3}(\delta\rho)^3+\frac{c}{4}(\delta\rho)^4 + \lambda_1\left(\Delta\delta\rho + q^2\delta\rho\right)^2.
\end{equation}
The first three terms with coefficients $a$, $b$ and $c$ are a Landau expansion of the free energy that sets the amplitude of layering. 
The last term enforces the constraint of equidistance layers with the wave number $q$. If we assume $\delta\rho = \cos(\mathbf{k}\cdot \mathbf{x})$, then $(\Delta\delta\rho + q^2\delta\rho)^2 = (-|\mathbf{k}|^2\delta\rho + q^2\delta\rho)^2$,
which is minimised when $|\mathbf{k}| = q$. The constant $\lambda_1$ is the weight coefficient of the equidistant layering penalty. The value of $\lambda_1$ scales approximately as $O(q^{-4})$ with units of $Nm^2$.
The wave number $q$ depends on the thickness of layer, i.e., $q = \frac{2\pi}{d_s}$ ($d_s = d_A$ or $d_C$). The layer thickness $d_s$ ranges from $nm$ to $\mu m$. The relation between the smA and smC layer thickness is usually $d_C = d_A\cos\theta_0$ (see Fig.\ref{fig:smAC}).
Exceptions are the so-called de Vries materials \cite{lagerwall2006current, de1977experimental}, the layer thickness of which is almost unchanged during the smA-smC transition. Note that the parameter $q$ is supposed to be proportional to $\frac{1}{\cos\theta_0}$. While, as the effect of $q$ has been studied in the smA model \cite{pevnyi-2014-article, xia-2021-article, xia-2023b-article}, and in order to focus on investigating the effect of the tilt angle $\theta_0$ in smC phases, we assume $q$ is a constant for both smA and smC in each case of the following numerical results. 
\\
(iii) The coupling term of $\Qvec$ and $\delta\rho$ is given by
\begin{equation}
    f_{tilt}(\delta\rho,\mathbf{Q}) = \lambda_2 \left(\mathrm{tr}\left(\mathcal{D}^2\delta\rho\left(\mathbf{Q}+\frac{\mathbf{I}_d}{d}\right)\right) + q^2\delta\rho\cos^2\theta_0\right)^2,
\end{equation}
which is responsible for maintaining $\theta_0$, the tilt angle between layer normal and the leading director of $\mathbf{Q}$ for smC phases, see Fig.~\ref{fig:smAC}.
Here, $\lambda_2$ represents the weight coefficient of the coupling effect, with a value approximately of $O(q^{-4})$ and units of $Nm^2$, $\mathcal{D}^2(\cdot)$ is the Hessian operator and $\mathbf{I}_d$ denotes the $d\times d$ identity matrix.
For the smA phase, we have $\theta_0 = 0$.
For the smC phase, the nonzero tilt angle $\theta_0$ is an essential characteristic of smC LCs and has a significant impact on their properties and behaviour.
Typical values of the tilt angle $\theta_0$ are in the range of $10^\circ\sim 30^\circ$ \cite{trittel2017smectic}. 
It can be controlled by temperature \cite{taylor-1970-article, urbanski2017liquid} and external forces such as electric fields \cite{garoff-1977-article, eremin-2008-article} and mechanical force \cite{trittel2017smectic, kramer-2011-article}.
If we assume $\delta\rho = \cos(q \mathbf{m}\cdot \mathbf{x})$, where $\mathbf{m}$, $\mathbf{x}$ represent the layer normal and spatial location respectively, we have $\mathcal{D}^2\delta\rho = -q^2 \delta\rho(\mathbf{m}\otimes\mathbf{m})$. Further, if $\Qvec$ is uniaxial with a leading director $\mathbf{n}$, i.e., $\Qvec = \mathbf{n}\otimes\mathbf{n} - \frac{\Ivec_d}{d}$, after calculation, the coupling term is minimised when $(\mathbf{m}\cdot\mathbf{n})^2 = \cos^2\theta_0$, i.e., the angle between the layer normal $\mathbf{m}$ and director $\mathbf{n}$ is $\theta_0$. 

By rescaling the free energy \eqref{eq:Functional} on $d$-dimensional domain $\Omega$ according to $\bar{\mathbf{x}} = \frac{\mathbf{x}}{\lambda}$, $\bar{F} = \frac{F\bar{K}}{K\lambda^{d-2}}$, $\bar{l} = \frac{l\lambda^2\bar{K}}{2K}$, $\bar{a} = \frac{\bar{K}a\lambda^2}{K}$, $\bar{b} = \frac{\bar{K}b\lambda^2}{K}$, $\bar{c} = \frac{\bar{K}c\lambda^2}{K}$, $\bar{q} = q\lambda$, $\bar{\lambda}_1 = \frac{\lambda_1\bar{K}}{K\lambda^2}$, $\bar{\lambda}_2 = \frac{\lambda_2\bar{K}}{K\lambda^2}$,
the non-dimensionalised total energy is then given by
\begin{align}
&\bar{F}(\delta\rho, \mathbf{Q})\nonumber\\
&= \int_{\Omega_0} -\frac{\bar{l}}{2} \text{tr}(\Qvec^2) - \frac{\bar{l}}{3} \text{tr}(\Qvec^3) + \frac{\bar{l}}{2} \left(\text{tr}(\Qvec^2)\right)^2+ \frac{\bar{K}}{2}|\nabla\Qvec|^2\nonumber\\
&+ \frac{\bar{a}}{2}(\delta\rho)^2+\frac{\bar{b}}{3}(\delta\rho)^3+\frac{\bar{c}}{4}(\delta\rho)^4 + \bar{\lambda}_1\left(\Delta\delta\rho + \bar{q}^2\delta\rho\right)^2\nonumber\\
&+ \bar{\lambda}_2 \left(\mathrm{tr}\left(\mathcal{D}^2\delta\rho\left(\mathbf{Q}+\frac{\mathbf{I}_d}{d}\right)\right) + \bar{q}^2\delta\rho\cos^2\theta_0\right)^2 \mathrm{d}\bar{\mathbf{x}}.
\label{eq:nondim}
\end{align}
Here, $\lambda^d$ represents a measure of domain size, $\Omega_0$ denotes the rescaled domain, and $\bar{K}$ stands for the non-dimensionalised elastic constant. The choice of $\bar{a}$, $\bar{b}$ and $\bar{c}$ simply follows \cite{pevnyi-2014-article, xia-2021-article}.
In the implementation, we omit the bars over the parameters and denote $\Omega_0$ as $\Omega$.
\textit{Results}--
We calculate the minimisers or the critical points of the free energy \eqref{eq:nondim} with respect to both $\mathbf{Q}$ and $\delta\rho$ or $\mathbf{Q}$ only in some smC cases, by employing finite element approximations to solve the corresponding Euler--Lagrange (EL) equations. 
Additionally, due to the nonlinearity in the EL equations, we utilise the deflated continuation algorithm, so to compute multiple solutions of the EL equations and the continuation of known solution branches. A more detailed description of the numerical methods, the stability calculation, and information on each simulation is provided in the Appendices.

\begin{figure}
    \includegraphics[width=1\columnwidth]{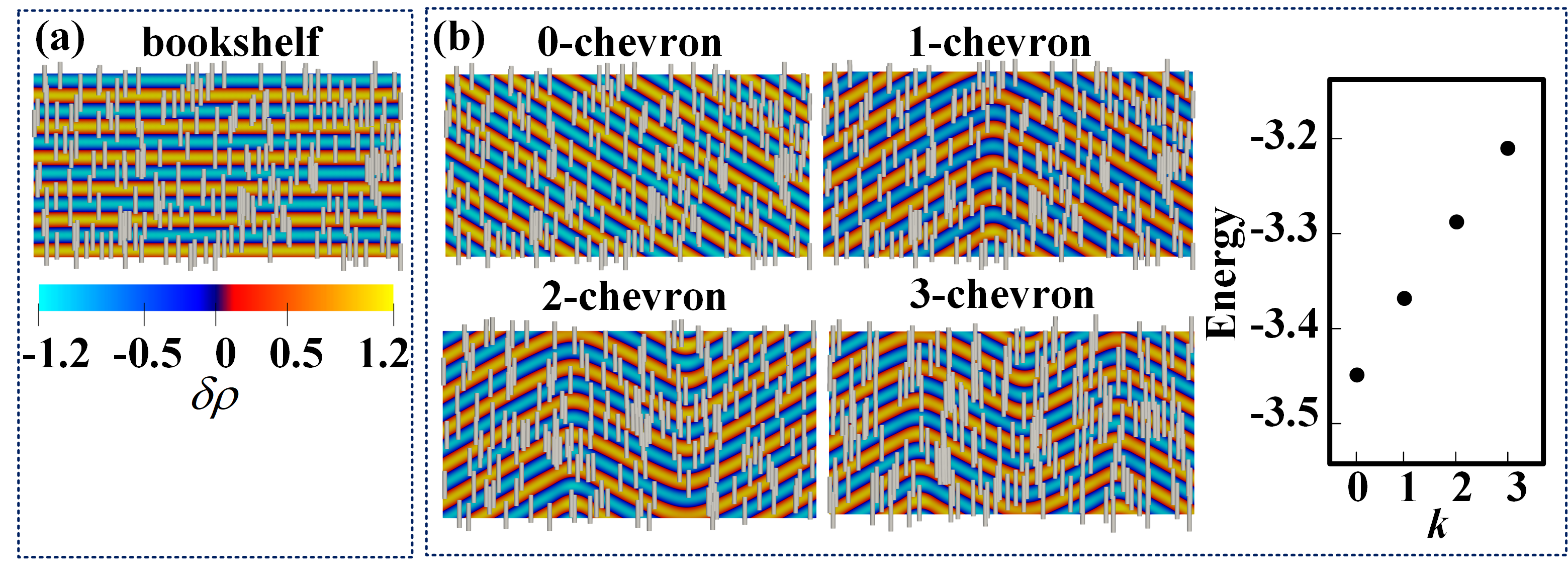}
    \caption{(a) The bookshelf structure in smA; (b) the $k$-chevron structures in smC with tilt angle $\theta_0=\frac{\pi}{6}$ and the plot of the free energy of $k$-chevron versus the $k$, number of foldings.
    Vertical directors are strongly imposed on top and bottom boundaries of the domain.
     The parameters are taken as $q=12\pi$ and $a=-10, b=0, c=10, K=l=0.3, \lambda_1=\lambda_2=10^{-5}$. The density variation $\delta\rho$ is represented by color from blue to yellow.
The director field $\mathbf{n}$ is the eigenvector associated with the largest eigenvalue of $\Qvec$ and visualised by uniform-length gray rods. We use the same visualisation method for Fig.~\ref{fig:dislocation} and Fig.~\ref{fig:pmhalf}(e-h). Note that directors are largely in vertical alignment due to the imposed boundary condition.
}
    \label{fig:chevron}
\end{figure}

\textit{(a) Bookshelf and chevron}--There are two related typical smectic structures: ``bookshelf" and ``chevron" \cite{biscari-2007-article,webster2003molecular}.
The imposition of equidistant smectic layer constraints results in the formation of horizontal layers, a characteristic feature of the smA phase. This structure is often referred to as a bookshelf configuration (Fig.~\ref{fig:chevron}(a)).
The corresponding bookshelf configuration for smC phases, the special case of a chevron structure with no folding layer ($0$-chevron), is shown in the first image of Fig.~\ref{fig:chevron}(b). This structure has been observed in numerous experiments by cooling a smA bookshelf or by the effect of confinement ~\cite{ricker-1987-article, lefort-2014-article,limat-1993-article}.

In Fig.~\ref{fig:chevron}(b), in each $k$-chevron structure with $k\geq 1$, the layers tilt with the tilt angle varying between $+\theta_0$ and $-\theta_0$, and bend with $k$ foldings. Given our assumption of the thin film limit, all molecules in our continuum model lie within the plane. The directors on the layers can tilt to either left or right by the angle $\theta_0$ and are predominantly vertical due to the structural symmetry. Near the chevron tips, the directors remain vertical, although the tilt angle between the layer normal and the director may not be well-maintained. Consequently, foldings are energetically unfavorable.
The free energy \eqref{eq:nondim} increases as $k$ increases, as observed in Fig.~\ref{fig:chevron}. In a 3D domain, the configuration near the chevron tips is discussed in \cite{limat-1995-article}. 
The $1$-chevron pattern in smC phases was often observed in experiments, as reported by \cite{ricker-1987-article, mckay-2004-article, jones-2015-article, biscari-2007-article}.
Although the configuration with fewer foldings is more energetically preferred, the $k$-chevron structure with multiple foldings is unavoidable in confinements like shells \cite{urbanski2017liquid} due to its special geometry. 

\begin{figure}
    \includegraphics[width=1\columnwidth]{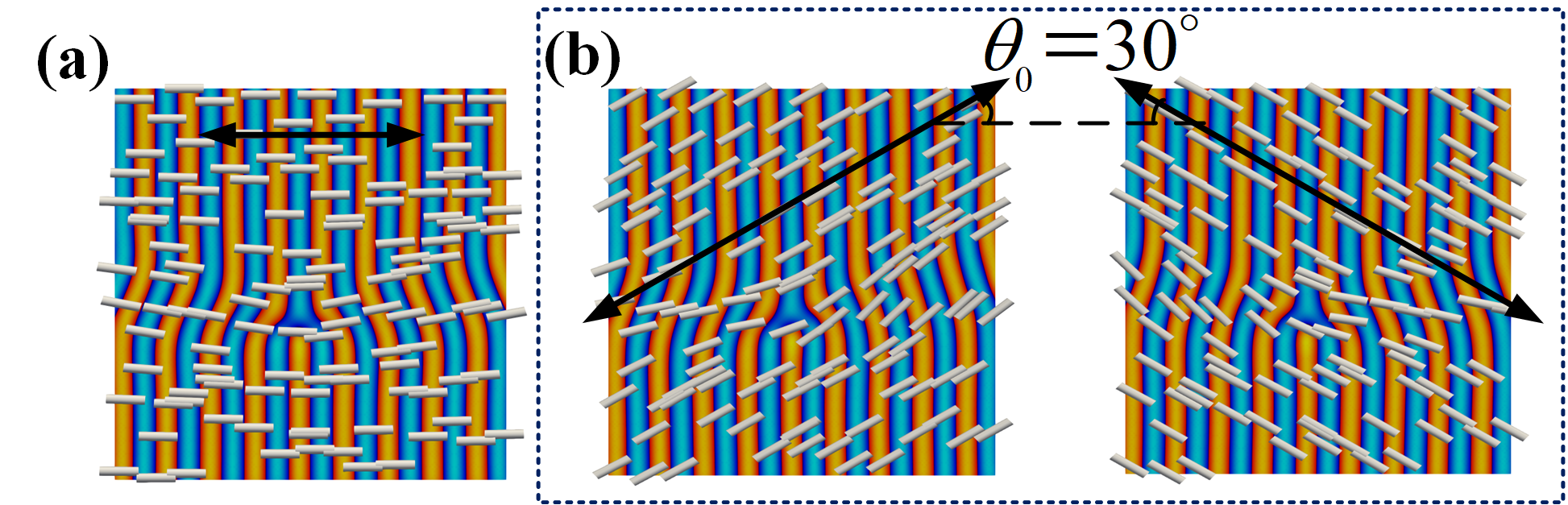}
    \caption{Dislocation structures in the (a) smA phase with horizontal directors strongly enforced at the top and bottom surfaces, and (b) smC phase with $\theta_0 = \frac{\pi}{6}$ with left- or right-tilted Dirichlet boundary conditions imposed on the top and bottom surfaces. 
    Black arrows represent the approximate main orientation of directors. Here, we take parameters $q=9\pi$, $a=-10$, $b=0$, $c=10$, $K=0.01$, $l=1$, $\lambda_1=\lambda_2=10^{-5}$.
    }
    \label{fig:dislocation}
\end{figure}

\textit{(b) Dislocation}--The edge dislocations are unique structures to materials with broken translational symmetry such as smectics \cite{meyer-1978-article, zhang-2015-article}.
By minimising the non-dimensionalised free energy \eqref{eq:nondim}  with an appropriate initial guess (refer to Scenario II in Appendix D), we recover a single dislocation profile in the smA phase, with two layers merging into one layer at the center of the domain as shown in Fig.~\ref{fig:dislocation}(a). 
Near the dislocation, the director changes along the layers and no longer conforms to the prescribed boundary conditions. The magnitude $|\delta\rho|$ of the density variation is weaker in regions where the director undergoes significant changes, as observed in the dark blue areas within light blue layers surrounding the dislocation, as shown in Fig.~\ref{fig:dislocation}. 
To examine the orientation variation in the smC phase around an analogous single dislocation within the domain, we use the layering pattern and director orientations from Fig.~\ref{fig:dislocation}(a) as the initial guess for the energy minimising procedure for smC phases. We apply appropriate left- and right-tilted orientational constraints on the directors at the top and bottom boundaries.
Fig.~\ref{fig:dislocation}(b) showcases two classes of orientations in each layer, naturally arising due to the symmetric tilting of the angle $\theta_0$ from the layer normal. 
Both profiles shown in Fig.~\ref{fig:dislocation}(b) exhibit nearly identical energy levels, differing by only approximately $10^{-3}$ in magnitude, underscoring the tilting symmetry of directors within each layer for smC.




\begin{figure}
\includegraphics[width=1\columnwidth]{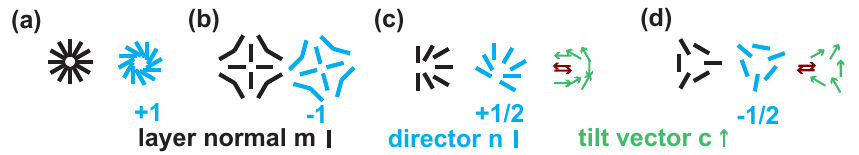}
    \includegraphics[width=1\columnwidth]{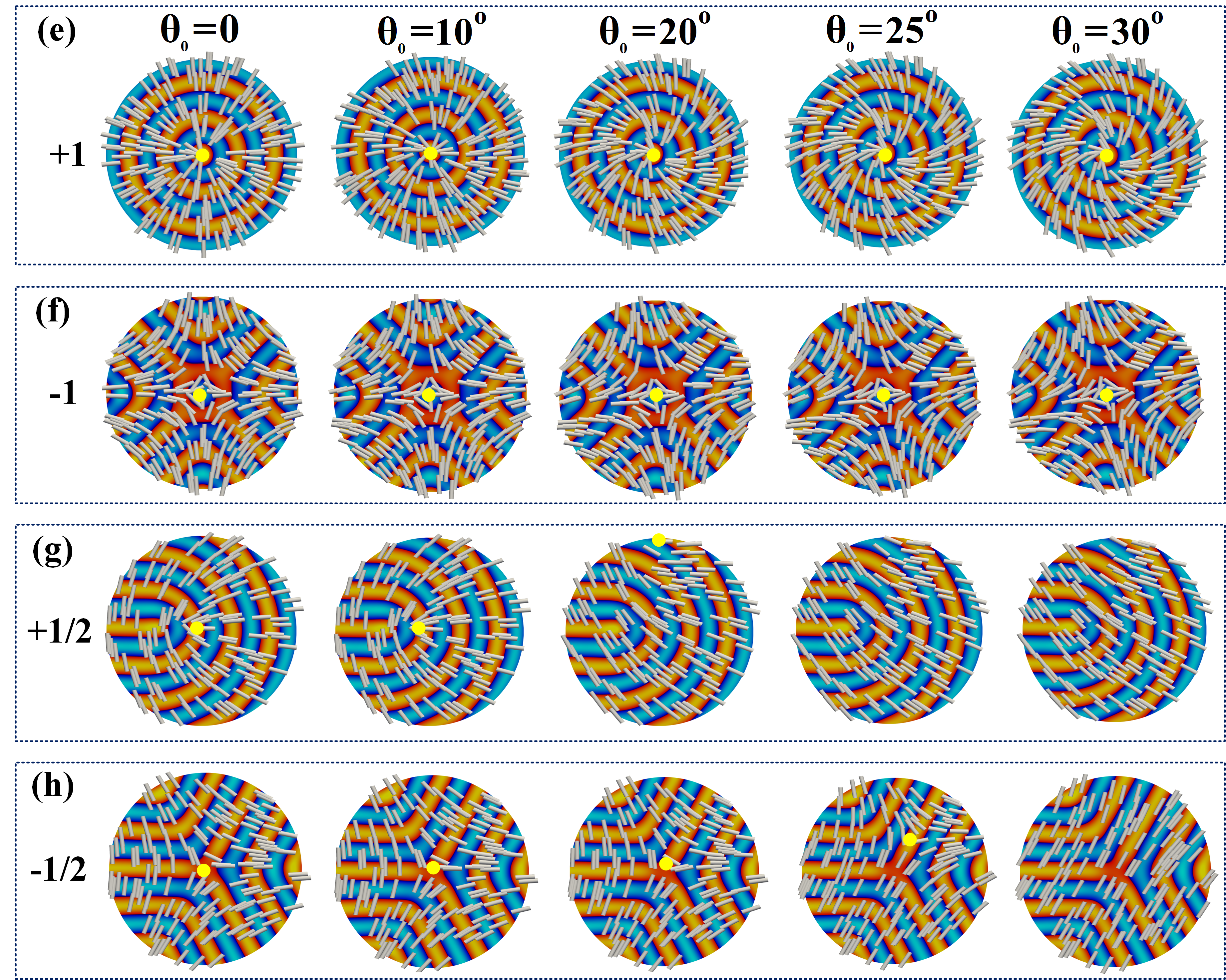}
    \caption{(a-d) The schematic illustration of expected $w$-charged defects ((a) $w = +1$, (b) $w = -1$, (c) $w = +1/2$, (d) $w = -1/2$) with the layer normal $\mathbf{m}=(\cos(w\vartheta), \sin(w\vartheta))$, director $\mathbf{n}=(\cos(w\vartheta+\theta_0), \sin(w\vartheta+\theta_0))$ with $\theta_0 = \pi/6$. 
 In (c-d), the tilt vector field is given by $\mathbf{c} = \frac{\mathbf{n}-(\mathbf{n}\cdot\mathbf{m})\mathbf{m}}{|\mathbf{n}-(\mathbf{n}\cdot\mathbf{m})\mathbf{m}|}$ and the discontinuity of $\mathbf{c}$ is marked by red arrows. (e-h) The $\pm1$, $\pm1/2$ defect structures in smA with $\theta_0=0$ and smC with various non-zero $\theta_0$. We set parameters as $q=10\pi$, $a=-10,b=0,c=10,K=0.3, l=5,\lambda_1=\lambda_2=10^{-5}$. Defects are marked with yellow dots.
    }
    \label{fig:pmhalf}
\end{figure}

\textit{(c) Defect}--The behaviour of defects in smC has been discussed intuitively in \cite{harth2020topological}. In the smA phase, the configuration near a $w$-charge defects, without loss of generality, can be described by the layer normal $\mathbf{m}=(\cos(w\vartheta), \sin(w\vartheta))$ and director $\mathbf{n}=(\cos(w\vartheta), \sin(w\vartheta))$, where $\vartheta$ is the azimuth angle on $xy$-plane. For smC phases, as we fix the layer structure, i.e., $\mathbf{m}=(\cos(w\vartheta), \sin(w\vartheta))$, if we only consider the tilt angle $\theta_0$, the expected director is $\mathbf{n}=(\cos(w\vartheta+\theta_0), \sin(w\vartheta+\theta_0))$ (see Fig.~\ref{fig:pmhalf}(a-d)). Due to the symmetry of the structure, here we only assume that the director tilts to the left relative to the layer normal. The defect charge of the expected director is still $w$. For $w\neq +1$, the addition of the phase angle $\theta_0$ in directors $\mathbf{n}$ is equivalent to rotating the configuration of layer normals $\mathbf{m}$ as a whole by $\frac{\theta_0}{(1-w)}$ (see Fig.~\ref{fig:pmhalf}(b-d)). For $w=+1$, the configurations of $\mathbf{m}$ and $\mathbf{n}$ are not rotationally degenerate, as shown in Fig.~\ref{fig:pmhalf}(a). The construction of $\mathbf{n}=(\cos(w\vartheta+\theta_0), \sin(w\vartheta+\theta_0))$ works well for integer-charged defects in the smC system. However, the half-integer charged defects are prohibited in the smC phase to some extent, due to the polar character of the tilt vector field $\mathbf{c} = \frac{\mathbf{n}-(\mathbf{n}\cdot\mathbf{m})\mathbf{m}}{|\mathbf{n}-(\mathbf{n}\cdot\mathbf{m})\mathbf{m}|} = (-\sin(w\vartheta), \cos(w\vartheta))$.
The tilt vector $\mathbf{c}$ represents the unit vector along the projection of $\mathbf{n}$ on the smectic plane. 
In contrast to the directors $\mathbf{n}$ and $-\mathbf{n}$, which are equivalent, and the layer normals $\mathbf{m}$ and $-\mathbf{m}$, which are also equivalent in both smC and smA phases, the tilt vector field $\mathbf{c}$ does not have ``director symmetry", i.e., $\mathbf{c}$ and $-\mathbf{c}$ are not equivalent \cite{lagerwall2004ferroelectric}.
In Fig.~\ref{fig:pmhalf}(c-d), the discontinuity of $\mathbf{c}$ is indicated by red arrows, signifying conflicting signs of $\mathbf{c}$.

We successfully observe the phenomena above in our numerical results in Fig.~\ref{fig:pmhalf}(e-h).
Firstly, we recover the profile with a $\pm 1$ or $\pm 1/2$ defect within the centre of the disc domain for the smA system. The smA state with $+1$ and $\pm 1/2$ are obtained by minimising the free energy \eqref{eq:nondim} with $\theta_0=0$ with certain specified boundary conditions on $\mathbf{Q}$ (refer to Scenario III in Appendix D). Strong surface anchoring is employed to induce the desired defects. Notably, the structure in the smA state with $-1$ defect, in fact, is a critical state rather than a local minimiser. 
In Fig.~\ref{fig:pmhalf}(f), near the $-1$ defect at the centre, four dislocations are observed where two layers merge into one. The magnitude $|\delta \rho|$ of density variation is lower at these dislocations, consistent with observations in Fig.~\ref{fig:dislocation}.
Then, we continue the tilt angle $\theta_0$ as the essential parameter in the deflated continuation algorithm (see Appendix B) on $\pm 1$ and $\pm\frac{1}{2}$ defect branches, respectively. To preserve energetically unfavourable defect structures and focus on the coupling between directors and layer normals, we maintain the pattern of $\delta\rho$ arising from smA experiments. We then minimise or calculate the critical point of the energy functional \eqref{eq:nondim} with respect to $\Qvec$ only.
As the tilt angle $\theta_0$ increases, the directors in Fig.~\ref{fig:pmhalf}(e) for $+1$ defect become spiral, the directors in Fig.~\ref{fig:pmhalf}(f) for $-1$ defect rotate as a whole by $\theta_0$. In Fig.~\ref{fig:pmhalf}(e-f), the $\pm 1$ defects remain at the centre, while in Fig.~\ref{fig:pmhalf}(g-h), the $\pm 1/2$ defects gradually move away from the centre point. Essentially, when the tilt angle $\theta_0$ is large enough, the $\pm 1/2$ defect disappears from the disc domain. 

\textit{(d) Toroidal focal conic domains (TFCDs)}--In a more complex scenario, we extend the domain into 3D, allowing for the emergence of experimentally observed defects known as TFCDs in smA LCs \cite{williams-1975-article}. A schematic description and experimental observation of confocal conics in smC are depicted in \cite{perez-1978-article, iida-2007-article}, but it has not been mathematically explored in prior research. In TFCDs, the smectic layers adopt a unique configuration, comprising stacked interior sections of tori, with a central line defect extending between the two substrates.

\begin{figure}
    \includegraphics[width=1\columnwidth]{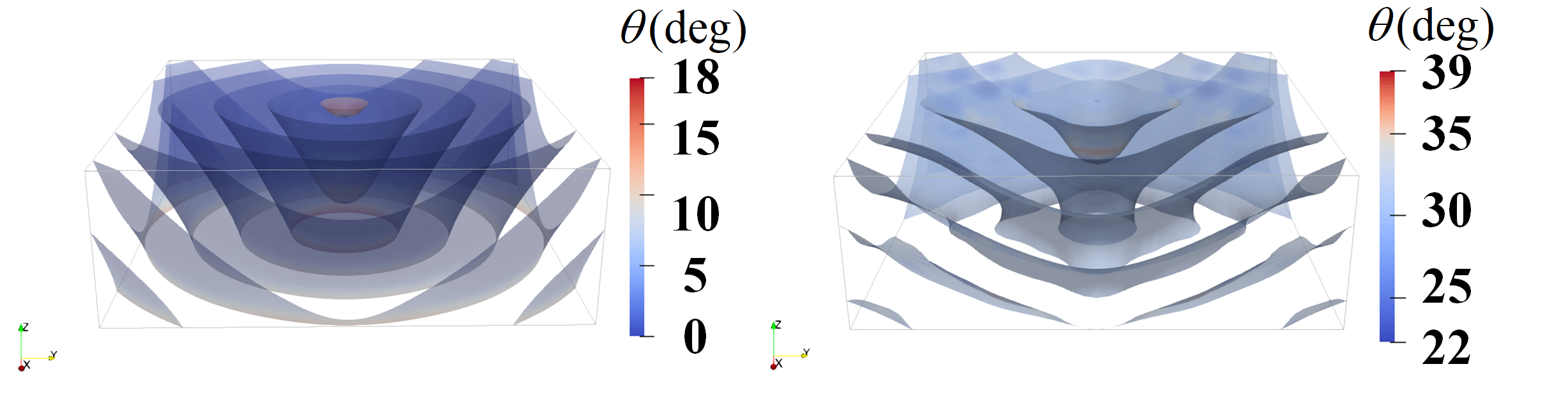}
    \caption{TFCD profiles in the scenarios of smA (left) with $\theta_0 = 0$ and smC  (right) with $\theta_0=\frac{\pi}{6}$. Here, parameters are taken as $q=10$, $a=-10, b=0, c=10, K=0.03, l=10, \lambda_1=\lambda_2=10^{-3}$. We plot the zero iso-surface of the density variation $\delta\rho$ to illustrate the layering structure, and colour them with the computed angle $\theta$ between directors and layer normals, where $\theta = \arccos{\frac{|\nabla \delta\rho \cdot \mathbf{n}|}{|\nabla \delta\rho|}}$, $\mathbf{n}$ is the unit eigenvector of $\mathbf{Q}$ corresponding to the largest eigenvalue.
    }
    \label{fig:tfcd}
\end{figure}

Our proposed model is effective in capturing the TFCD structure in both smA and smC, as shown in Fig.~\ref{fig:tfcd}. 
In the smA case, it is expected that the angle between the directors and the layer normals $\theta$ should be close to $\theta_0 = 0$, as verified in the left panel of Fig.~\ref{fig:tfcd}. Notably, non-zero values of $\theta$ tend to concentrate around the central axis of the computational domain and the bottom surface. This concentration is a result of the presence of a defect line in TFCDs, which coincides with the central axis, and the application of radial confinements to the directors on the bottom surface.
It is noted that directors near the bottom surface deviates from layer normals, this is actually due to the surface anchoring predominates over TFCD structures near the surface.
For the smC case with a specific value of $\theta_0=\frac{\pi}{6}$, the calculated values of $\theta$ exhibit a consistent variation around $\theta_0 = \frac{\pi}{6}$ in the right panel of Fig.~\ref{fig:tfcd}. Similar to the smA phase, deviations from $\frac{\pi}{6}$ tend to be concentrated around the central axis and the bottom surface. This similarity in behaviour between the two phases highlights the effectiveness of the proposed tensorial model in describing both smA and smC phases within a unified framework.

\textit{Conclusions and Discussions}--In this letter, we proposed a continuum mathematical model for smC (and smA) phases with a real-valued density variation for layering structures and a tensor-valued order parameter for orientations of LC molecules. This model explicitly includes the tilt 
angle $\theta_0$ between directors and layer normals, which serves as a crucial model parameter to distinguish between smA and smC phases.
To enhance the reliability and applicability of our model, we performed a series of 2D and 3D numerical experiments to recover
typical structures in smA and smC phases, such as bookshelves and chevrons, dislocations, defects and TFCDs.

 
Prospective future research includes the exploration of phase transitions \cite{freelon-2011-article, biscari-2007-article}, as our model provides a unified framework that can describe the first- or second-order phase transitions between isotropic, nematic, smA, and smC phases.
This new tensorial model can serve as a foundation for investigating the complex dynamics of topological defects and dislocation in layered smectics, which have an additional positional order compared to nematics. 

\textit{Acknowledgments}--The work of J.X. is supported by the National Natural Science Foundation of China (Grant No.~12201636), Research Fund of National University of Defense Technology (Grant No.~ZK22-37), Science and Technology Innovation Program of Hunan Province (Grant No.~2023RC3013), Hunan Provincial Natural Science Fund for Excellent Youths (Grant No.~2023JJ20046), and Young Elite Scientists Sponsorship Program by CAST (Grant No.~2023-JCJQ-QT-049). Y.H. is supported by a Leverhulme Project Research Grant RPG-2021-401.



\setcounter{figure}{0}
\setcounter{section}{0}
\section*{Appendix A: Finite Element Methods}\label{ap:FEM}

Minimising the free energy (7) of the main text presents significant challenges in numerics due to the presence of the Hessian term, which necessitates, from analytical viewpoints, that $\delta \rho \in \mathcal{H}^2\subset \mathcal{C}^1$.
Considerable attention is required in determining suitable finite element approximations, with a focus on both conforming and nonconforming aspects. For a comprehensive introduction to finite element methods, refer to works such as \cite{ciarlet-fembook, brenner-book, brenner-2011-book}.
For simplicity, we follow a strategy used in our previous work \cite{xia-2021-article} to deal with this similar challenge. To this end, we apply the so-called $\mathcal{C}^0$ {interior penalty} approach \cite{brenner-2011-book}, resulting in nonconforming discretisations.
In essence, this approach employs only $\mathcal{C}^0$ elements, meaning that the approximations are continuous without necessarily having continuous first derivatives. To overcome the non-conformity issue, inter-element jumps in the first derivatives are penalised, effectively enforcing $\mathcal{C}^1$-conformity in a weak sense.

Hence, we add an inter-element penalty term to the energy functional (7) of the main text in all of our implementations, leading to
\begin{equation}
    F_\gamma(\delta \rho, \Qvec) := F(\delta \rho, \Qvec) + \sum_{e\in \mathcal{E}_I} \int_{e} \frac{\gamma}{2h_{e}^3} \left( \llbracket \nabla \delta\rho \rrbracket \right)^2.
\end{equation}
Here, $\gamma$ is the penalty parameter (we fix $\gamma=1$ throughout this work to ensure the convergence as discussed in \cite{xia-2023-article}), $\mathcal{E}_I$ corresponds to the collection of interior facets, which are edges or faces within a mesh, $h_{e}$ denotes the size of an edge or face $e$, and the jump operator applied to a vector $\nabla \delta\rho$ on a facet $e$ shared by two adjacent cells, labelled as $K_-$ and $K_+$, is defined as $\llbracket \nabla \delta\rho \rrbracket = (\nabla \delta\rho)_-\cdot \nu_- + (\nabla \delta\rho)_+ \cdot \nu_+$, where $\nu_-$ and $\nu_+$ denote the restriction of the outward normal to $K_-$ and $K_+$, respectively.
It is noteworthy that the numerical analysis of this choice of discretisation has been reported in detail in \cite{xia-2023-article} for a similar minimisation problem modelling smA LCs.
Utilising the $\mathcal{C}^0$ interior penalty method offers several advantages, combining both convenience (no extra construction of sophisticated discretisations) and efficiency (since the weak form remains straightforward, with only minor adjustments compared to a conforming method) in computations. Moreover, this approach employs fewer degrees of freedom in comparison to a fully discontinuous method, e.g., Discontinuous Galerkin methods \cite{suli-2007-article}.

Specifically, in two dimensions, we employ quadrilateral meshes where the function space $\mathbb{CG}_k$ is defined through tensor products of polynomials with degrees up to $k$ in each coordinate direction. This results in spaces of piecewise biquadratic functions for $\mathbb{CG}_2$ and piecewise bicubic functions for $\mathbb{CG}_3$. For the tensor $\Qvec$, we impose the constraint of it being a symmetric and traceless tensor, which gives rise to two independent components in two dimensions. Consequently, we seek the components of $\Qvec$ in $\mathbb{CG}_2^2$, while representing $\delta\rho$ in $\mathbb{CG}_3$.
In three dimensions, we utilise hexahedral meshes with similar tensor-product spaces, resulting in spaces of piecewise triquadratic functions for $\mathbb{CG}_2$ and piecewise tricubic functions for $\mathbb{CG}_3$. In this context, the tensor $\Qvec$ possesses five independent components. Therefore, we seek its components in $\mathbb{CG}_2^5$, while retaining $\delta\rho$ in $\mathbb{CG}_3$.

As the total energy (1) under consideration is nonlinear in nature, we employ Newton's method with an $L^2$ linesearch, as outlined in \cite[Algorithm 2]{brune2015}, as the outer nonlinear solver. The nonlinear solver is considered to have converged when the Euclidean norm of the residual falls below $10^{-8}$ or decreases by a factor of $10^{-14}$ from its initial value, whichever occurs first.
For the inner solves, the linearised systems are tackled using the sparse LU factorization library MUMPS \cite{mumps}. This solver framework is implemented within the Firedrake library \cite{firedrake}, which relies on PETSc \cite{petsc} for solving the resulting linear systems.
In the context of the $\mathcal{C}^0$ interior penalty approach, the mesh scale $h_e$ is selected as the average of the diameters of the cells located on either side of an edge or face.

\section*{Appendix B: The deflated continuation algorithm}\label{ap:deflate}

In addressing a parameter-dependent nonlinear problem, such as seeking solutions to the Euler--Lagrange equation for the proposed energy functional (7) of the main text, where the tilt angle $\theta_0$ serves as the parameter, it is inherent to encounter multiple solutions. This diversity offers a comprehensive insight into the potential states and configurations attainable by the system under specified conditions.
In this manuscript, we adopt the \textit{deflated continuation algorithm} to systematically identify multiple solutions. For a more detailed exploration of this algorithm, one may refer to \cite{farrell-birkisson-2015-article}. However, we present a self-contained explanation of the algorithm, as also elucidated in \cite{xia-2021-article}.

The algorithm seamlessly integrates two essential ingredients: \textit{deflation} and \textit{continuation}, where the latter is a widely utilised technique for challenging nonlinear problems to facilitate convergence. We use standard continuation \cite{seydel2010} in the implementation of this work. Hence, we briefly introduce the idea of deflation and give an overview of the algorithm below.

Consider a general parameter-dependent nonlinear problem 
\begin{equation}\label{problem}
    f(u,\lambda) = 0\quad \text{for }u\in U\ \text{and } \lambda\in [\lambda_\mathrm{min},\lambda_\mathrm{max}],
\end{equation}
where $U$ is an admissible space for $u$ and $\lambda$ is the parameter. In our context, $f$ is the residual of the Euler--Lagrange equation for the proposed energy functional Eq.~(7) of the main text, $u$ represents the variable pair $(\delta \rho, \Qvec)$, and $\lambda$ represents the tilt angle $\theta_0$.

For a fixed parameter $\lambda^{\star}$, problem \eqref{problem} then becomes
\begin{equation}\label{fix-problem}
    G(u)\coloneqq f(u,\lambda^{\star}) = 0.
\end{equation}
Initially, we employ the classical Newton iteration, starting from an initial guess denoted as $u^0$, in order to find a solution, referred to as $u^{\star}$, for the equation \eqref{fix-problem}. Following this, we proceed with a process called \emph{deflation} for this identified solution.

The primary objective of deflation is to formulate a new nonlinear problem denoted as $H(u)$, which shares the same solutions as the original problem $G$, except for the now-known solution $u^{\star}$. Subsequently, we can apply Newton's method to this new problem $H$, once again commencing from the initial guess $u^0$. If this iterative process converges, it will lead to the identification of a different solution. It is important to highlight that the deflation technique effectively provides only a single initial guess for the initial parameter value in our implementation.

In this work, we construct the deflated problem $H$ via
\begin{equation*}
    H(u) \coloneqq \left(\frac{1}{\|u-u^{\star}\|^2} + 1\right)G(u) = 0,
\end{equation*}
where the norm used in this work is
\begin{equation*}
\|u\|^2 = \|(\delta \rho, \Qvec)\|^2 = \int_\Omega \left( (\delta \rho)^2 + \sum_i \Qvec_i^2 \right)\,
\end{equation*}
with $\Qvec_i$ being the $i^\mathrm{th}$ component of the vector proxy for $\Qvec$ (i.e., 2 components in two dimensions, 5 in three dimensions).
Under mild assumptions, it can be proven that Newton's method applied to $H$ will not converge to $u^{\star}$ again.

Having introduced the concept of deflation for a single nonlinear problem, we now provide a concise overview of the deflated continuation algorithm.
During a continuation step from $\lambda^-$ to $\lambda^+$, assuming that $m$ solutions $u_1^-, u_2^-, \dots, u_m^-$ are already known at $\lambda^-$, the next step unfolds in two distinct phases.
In the initial phase, each solution $u_i^-$ is continued from $\lambda^-$ to $\lambda^+$, resulting in corresponding solutions denoted as $u_i^+$.
Concurrently, as each solution $u_i^+$ is computed, it is deflated, effectively removing it from consideration in the nonlinear problem at $\lambda^+$.
Subsequently, the search phase of the algorithm initiates. Each previously known solution $u_i^-$ is reintroduced as an initial guess for the nonlinear problem at $\lambda^+$.
As discussed earlier, the deflation mechanism ensures that the Newton iteration will not converge to any of the previously identified solutions $u_i^+$. Therefore, if Newton's method achieves convergence, it signifies the discovery of a new, hitherto unknown solution.
In the event that an initial guess leads to the identification of a new solution, it undergoes deflation, and the initial guess is repeatedly utilised until it no longer succeeds. Once all initial guesses from $\lambda^-$ have been exhausted, the current continuation step concludes, and the algorithm progresses to the subsequent step.
These aforementioned phases are reiterated for each continuation step until $\lambda$ attains the desired target value of $\lambda_\mathrm{max}$.

In the implementation results present in the main text, we provide further detail on the utilisation of both the deflation and continuation components of the algorithm. In cases where a series of values for parameter $\theta_0$ are considered, such as $\theta_0$ ranging from $10^\circ$ to $30^\circ$ as depicted in Fig.~4 of the main text, the algorithm predominantly employs the continuation part. 
In scenarios where multiple solution profiles emerge for a single value of parameter $\theta_0$, the deflation component of the algorithm plays a significant role in identifying these distinct solutions.

\section*{Appendix C: Stability calculation}
We focus on minimising our proposed free energy (7) of the main text for smectic LCs. These minimising profiles inherently possess lower energy and align with the underlying physics. Consequently, evaluating the stability of the stationary solutions obtained during the minimisation process becomes a crucial step.
To assess the stability of each solution profile, we employ Cholesky factorisation to compute the inertia of the Hessian matrix associated with the energy functional, as implemented in MUMPS \cite{mumps}. We classify a solution as \textit{minimiser} if its Hessian matrix is positive semidefinite. Conversely, the presence of any non-zero number of negative eigenvalues in the Hessian matrix characterises a solution as \textit{non-minimising critical point} \cite{num-op99}.
It is noteworthy that in all the solution profiles presented in this manuscript, zero eigenvalues were not observed in the Hessian matrices. 
Furthermore, all configurations explored in this study are stable, except for the $-1$ defect profiles in Fig.~4(f) of the main text. 
\section*{Appendix D: Details on each simulation}\label{ap:details}
\subsection*{Scenario I: bookshelf and chevron}
In this straightforward example, we examine a rectangular region covering $\Omega = [-1,1]\times [0,1]$. The domain is discretised into a grid of $60\times 30$ quadrilaterals, each further crossly subdivided into four triangular elements. Along top $\Gamma_t$ and bottom $\Gamma_b$ boundaries of the rectangle $\Omega$, vertically aligned directors are enforced for both smA and smC cases, i.e., $\Qvec|_{\Gamma_t\bigcup\Gamma_b} = \Qvec_{v}=\begin{bmatrix}
    -1/2 & 0\\
    0 & 1/2
\end{bmatrix}$.
We take the following initial guess for both smA case as present in Fig.~2(a) and smC case as in Fig.~2(b) of the main text:
\begin{equation*}
\delta\rho_{0} = \cos(q y),\quad \Qvec_{0} = \Qvec_v.
\end{equation*}
We start from this initial guess and use the deflation method to find multiple $k$-chevron states in Fig.2(b).
The parameters are consistently taken as $q=12\pi$ and $a=-10, b=0, c=10, K=l=0.3, \lambda_1=\lambda_2=10^{-5}$.


It is noteworthy that multiple locally stable solutions exist as minimisers of the free energy under the conditions described in this subsection. Upon examination, solutions within the same class are found to maintain similar energy levels. For instance, in addition to the present left-tilted 0-chevron profile of smectic layers as shown in Fig.2(b) of the main text, there also exists a right-tilting 0-chevron configuration. The energy difference between these configurations is only on the order of $10^{-4}$.

\subsection*{Scenario II: dislocation}

In this scenario, we examine the domain $\Omega = [0,2]\times [0,2]$, which is uniformly discretised into a grid of $30\times 30$ squares, each further subdivided into four triangular elements. We impose strong boundary conditions on the director at the top surface $\Gamma_t=\{y=2\}$ and bottom surface $\Gamma_b=\{y=0\}$, while keeping the left and right lateral surfaces free with prescribed constraints. For the smA case, horizontal directors are enforced, i.e., $\Qvec|_{\Gamma_t\bigcup\Gamma_b} = \Qvec_{h} =\begin{bmatrix}
    1/2 & 0\\
    0 & -1/2
\end{bmatrix}$.
Meanwhile, the following initial guess of variables $\delta\rho$ and $\Qvec$ is taken: 
\begin{equation*}
\delta\rho_{0} = \begin{cases}
        \cos(9 \pi x) & \text{if } y>1\\
        \cos(10 \pi x) & \text{if } y\le 1
    \end{cases},\quad
\Qvec_{0} = \Qvec_h.
\end{equation*}
As for the smC case present in Fig.~3(b) of the main text, we simply use the obtained smA solution (i.e., Fig.~3(a)) as the initial guess for the smC model with $\theta_0=\pi/6$. Here, due to the natural tilting symmetry of smC molecules, left- and right-tilted directors are constrained on top and bottom surfaces.
We specify the values of parameters throughout this scenario as $q=9\pi$ and $a=-10, b=0, c=10, K=0.01, l = 10, \lambda_1=\lambda_2=10^{-5}$.

\subsection*{Scenario III: defect}
In this scenario, we opt for a circular domain $\Omega$ with unit radius from the origin, with model parameters taking $q=10\pi$, $a=-10, b=0, c=10, K=0.3, l=5, \lambda_1=\lambda_2=10^{-5}$.

For smA cases of defects with $w=+1, \pm 1/2$ in Fig. 4(e,g-h) of the main text, we impose Dirichlet boundary conditions for the orientations:
\begin{equation*}
\begin{aligned}
    &\Qvec|_{\partial\Omega} = \Qvec_b =
    \begin{bmatrix}
        \cos^2(\alpha) - \frac{1}{2} & \cos(\alpha)\sin(\alpha)\\
        \cos(\alpha)\sin(\alpha) & \sin^2(\alpha)-\frac{1}{2}
    \end{bmatrix},
    \end{aligned}
\end{equation*}
with $\alpha=w\vartheta$, where $\vartheta= \arctan(\frac{y}{x})$. The initial guess of $\delta\rho$ and $\Qvec$ is given by
\begin{equation*}
    \delta\rho_{ic} = \cos\left(q (\cos(\alpha)x+\sin(\alpha)y)\right), \quad
    \Qvec_{ic} = \Qvec_b.
\end{equation*}
As for the smC cases of defects with $w=+1, \pm 1/2$, we fix the layering pattern from the obtained smA profile and minimise only with respect to variable $\Qvec$ without boundary conditions of $\Qvec$. We use the continuation method to find minimisers with $\theta_0$ taking the value of $0, 10^\circ, 20^\circ, 25^\circ$ and $30^\circ$ respectively.
Note that in smC case of $+1$ defect, all present profiles are minimisers to the free energy (7) of the main text.

Due to the unavoidable appearance of dislocation structures,
for the $w = -1$ case in Fig.~4(f) of the main text, it is hard to find smA or smC minimisers with $-1$ defects.
The initial guess for smA profiles (i.e., the left panel) is set as
\begin{equation*}
    \delta\rho_{ic} = \cos\left(8 \pi (\cos({\alpha}) x+\sin({\alpha}) y)\right), \quad
    \Qvec_{ic} = \Qvec_b.
\end{equation*}
We found the non-minimising critical state with a $-1$ defect, as shown in the left panel of Fig.~4(f) in the main text.
Again, to obtain the corresponding smC profiles, we fix the layering pattern from the obtained smA profile and minimise only with respect to variable $\Qvec$, but this time with a stronger constraint, Dirichlet boundary enforcement 
\begin{equation*}
\begin{aligned}
    &\Qvec|_{\partial\Omega}=
    \begin{bmatrix}
        \cos^2(\alpha+\theta_0) - \frac{1}{2} & \cos(\alpha+\theta_0)\sin(\alpha+\theta_0)\\
        \cos(\alpha+\theta_0)\sin(\alpha+\theta_0) & \sin^2(\alpha+\theta_0)-\frac{1}{2}
    \end{bmatrix},
    \end{aligned}
\end{equation*}
with $\alpha=-\vartheta$.

\subsection*{Scenario IV: toroidal focal conic domains}

We discretise a box $\Omega = [-1.5,1.5]\times [-1.5,1.5]\times [0,1]$ into $6\times 6\times 5$ hexahedra, and label the boundary faces of $\Omega$ as
\begin{align*}
    &\Gamma_{left} = \{(x,y,z): x=-1.5\}, \, \Gamma_{right} = \{(x,y,z): x=1.5\},\\
    &\Gamma_{back} = \{(x,y,z): y=-1.5\}, \, \Gamma_{front} = \{(x,y,z): y=1.5\},\\
    &\Gamma_{bottom} = \{(x,y,z): z=0\}, \, \Gamma_{top} = \{(x,y,z): z=1\}.
\end{align*}

In the smA configurations, illustrated in Fig.~5 of the main text, we adhere to the formulation of boundary conditions and initial guesses for the smA case as elaborated in the Supplemental Material of \cite{xia-2021-article}.
Specifically, we consider the following surface energy
\begin{equation}
    \int_{\Gamma_{bottom}} \frac{W}{2}\left| \Qvec-\Qvec_{radial}\right|^2
    + \int_{\Gamma_{top}} \frac{W}{2}\left| \Qvec-\Qvec_{vertical} \right|^2,
\end{equation}
where $W=1$ denotes the weak anchoring weight,
\begin{equation*}
    \Qvec_{radial} = \begin{bmatrix}
    \frac{x^2}{{x^2+y^2}}-\frac{1}{3} & \frac{xy}{{x^2+y^2}} & 0\\
    \frac{xy}{{x^2+y^2}} & \frac{y^2}{{x^2+y^2}}-\frac{1}{3} & 0\\
    0 & 0 & -\frac{1}{3}
\end{bmatrix}
\end{equation*}
represents an in-plane ($x$-$y$ plane) radial configuration of the director, and
\begin{equation*}
    \Qvec_{vertical} = \begin{bmatrix}
    -\frac{1}{3} & 0 & 0\\
    0 & -\frac{1}{3} & 0\\
    0 & 0 & \frac{2}{3}
\end{bmatrix}
\end{equation*}
gives a vertical (i.e., along the $z$-axis) alignment configuration of the director.
Furthermore, we take the initial guess of $\delta\rho$ and $\Qvec$:
\begin{equation*}
    \delta\rho_{ic}^A = \cos(3\pi z), \quad \Qvec_{ic}^A=\mathbf{n}_{ic}\otimes\mathbf{n}_{ic}-\frac{\Ivec_3}{3},
\end{equation*}
where
\begin{equation*}
    \mathbf{n}^A_{ic} \coloneqq \begin{bmatrix} n_1\\ n_2\\ n_3\end{bmatrix} = \frac{1}{m}
        \begin{bmatrix}
             x\left(\sqrt{x^2+y^2} - R\right)\\
             y\left(\sqrt{x^2+y^2}-R\right)\\
             z\left(\sqrt{x^2+y^2}\right)
        \end{bmatrix},
\end{equation*}
and
\begin{equation*}
    m=\sqrt{x^2+y^2}\sqrt{\left(R-\sqrt{x^2+y^2}\right)^2+z^2}.
\end{equation*}
Here, the initial guess for the $\Qvec$-tensor is calculated based on the mathematical representation for a family of tori, and we have chosen a major radius $R=1.5$ in our implementation.

Regarding the smC configuration, we formulate the initial guess for the $\Qvec$-variable by leveraging the initial guess from the smA configuration. The boundary conditions of $\Qvec$ are rigorously enforced with the identical choice of the initial guess.
More specifically,
\begin{equation*}
     \Qvec_{ic} = \begin{cases}
        \mathbf{n}_{tilt}\otimes \mathbf{n}_{tilt} - \frac{\Ivec_3}{3} & \text{if } x^2 + y^2 \ge R^2,\\
        \mathbf{n}_{rotate}\otimes \mathbf{n}_{rotate} - \frac{\Ivec_3}{3} & \text{if } x^2 + y^2 < R^2,
    \end{cases}
\end{equation*}
where
\begin{equation*}
    \mathbf{n}_{tilt} = \begin{bmatrix}
    -\mathrm{sign}(x) \left|\cos\left(\arctan(\frac{y}{x})\right)\right| \sin(\theta_0)\\
    -\mathrm{sign}(y) \left|\sin\left(\arctan(\frac{y}{x})\right)\right| \sin(\theta_0)\\
\cos(\theta_0)
\end{bmatrix},
\end{equation*}
\begin{equation*}
\mathbf{n}_{rotate} = \begin{bmatrix}
\sin\left(\arccos(n_3)+\theta_0\right) \cos\left(\arctan(\frac{n_2}{n_1})\right)\\
\sin\left(\arccos(n_3)+\theta_0\right) \sin\left(\arctan(\frac{n_2}{n_1})\right)\\
\cos\left(\arccos(n_3)+\theta_0\right)
\end{bmatrix}.
\end{equation*}
It is important to note that $\mathbf{n}_{tilt}$ is derived from the schematic illustration in \cite[Fig. 2]{perez-1978-article}, while $\mathbf{n}_{rotate}$ effectively rotates $\mathbf{n}^A_{ic}$ azimuthally by an angle of $\theta_0$.
Throughout this scenario, parameters are taken as $q=10$, $a=-10, b=0, c=10, K=0.03, l=10, \lambda_1=\lambda_2=10^{-3}$.
\bibliography{ref}
\end{document}